\documentclass[reprint,prx,superscriptaddress,showkeys,longbibliography]{revtex4-2}

\usepackage{amsmath,amssymb,siunitx,physics,graphicx}
\usepackage[final]{changes}
\begin{document}


\title{Synthetic area spread in two-dimensional arrays of superconducting quantum interference filters}

\author{R. D. Monaghan}
\affiliation{Quantum and Nano Technology Group (QuaNTeG), School of Chemical Engineering, Adelaide University, North Terrace Campus, Adelaide, 5005, South Australia, Australia}

\author{J. L. Marenkovic}
\affiliation{Quantum and Nano Technology Group (QuaNTeG), School of Chemical Engineering, Adelaide University, North Terrace Campus, Adelaide, 5005, South Australia, Australia}

\author{G. C. Tettamanzi}
\email[]{giuseppe.tettamanzi@adelaide.edu.au}
\affiliation{Quantum and Nano Technology Group (QuaNTeG), School of Chemical Engineering, Adelaide University, North Terrace Campus, Adelaide, 5005, South Australia, Australia}

\date{\today}

\begin{abstract}

    Superconducting Quantum Interference Devices (SQUIDs), formed by incorporating Josephson junctions into loops of superconducting material, are the backbone of many modern quantum sensing systems. It has been demonstrated that, by combining multiple SQUID loops into a two-dimensional (2D) array, it is possible to fabricate ultra-high-performing radio-frequency sensors. However, to function as absolute magnetometers, current-in-use arrays require the area of each SQUID loop in the array to be incommensurate. In doing so forbids the achievement of their full potential of performance, limited only by the standard quantum limit. This is because imposing incommensurability in the areas contrasts with optimised performance in each single SQUID loop. In this work, we report that by selectively inserting "bare” sections of a superconducting circuit with no Josephson junctions, 2D SQUID arrays can operate as an absolute magnetometer even when no physical area spread is applied. Based on a generalisation of \added{currently} available theories, a complete analytical formulation for the one-to-one correspondence between the distribution of these bare loops and what we call a ”synthetic areas spread” is unveiled. This synthetic spread represents the equivalent physical spread of incommensurate SQUID loops that you would use to obtain the absolute Voltage-Magnetic Flux response if no bare loops were in use. Our work opens the way to a broader use of this technology for the fabrication of ultra-high-performance absolute quantum sensors. Our approach is also experimentally verified by fabricating several 2D Superconducting Quantum Interference Filter (SQIF) arrays incorporating bare superconducting loops and by demonstrating that they behave in alignment with what is suggested by our theory.
    
\end{abstract}


\maketitle

\section{Introduction} 
\label{sec::Intro}

The building block of superconducting based magnetometers is the superconducting quantum interference device (SQUID). In its most conventional implementation, i.e. the DC SQUID, this technology consists solely of a loop of superconducting material containing two Josephson junctions; a Josephson junction simply being a small barrier separating two superconducting regions~\cite{tinkhamIntroductionSuperconductivity1975}. This technology has matured to several commercial applications, such as the ones finalised for the operations of Magneto-Encephalography (MEG) machines~\cite{weinstockSQUIDSensorsFundamentals1996}. However, the Voltage-Magnetic Flux (VMF) response in conventional DC SQUIDs is highly noise sensitive and non-linear, making them impractical for some applications~\cite{tinkhamIntroductionSuperconductivity1975,oppenlanderTwoDimensionalSuperconducting2003,mitchell2DSQIFArrays2016,kornevSQIF2015}, especially when bandwidths in the Radio Frequencies (RF) ranges are needed. {Because of this, as fabrication techniques improve, interest has turned to two-dimensional arrays of SQUID loops that, by integrating a large number of SQUID cells in a serial and parallel electrical configuration, promise greater device sensitivity~\footnote{Sensitivity in a SQIF is sometimes more specifically associated in the literature to the concept of transfer function~$\frac{dV}{dB_{Ext}}$ or to Noise Sensitivity sometimes expressed in~$\frac{pT}{\sqrt{Hz}}$, we are here just mentioning sensitivity in its broader sense which may include both these concepts.}. When appropriately designed as a Superconducting Quantum Interference Filter (SQIF), the ability to function as absolute magnetometers is gained~\cite{oppenlanderTwoDimensionalSuperconducting2003,mitchell2DSQIFArrays2016,kornevSQIF2015}.}

Since it was theoretically proposed~\cite{Haussler2001,oppenlanderNonEnsuremathPhi2000} and experimentally demonstrated~\cite{oppenlanderTwoDimensionalSuperconducting2003,oppenlanderSuperconductingMultipleLoop2001,oppenlanderHighlySensitiveMagnetometers2002} the SQIF concept has been investigated and exploited by a continuously growing number of groups with respect to its basic properties\cite{Kornev2004,Oppenlaender2005,Kornev2009,Kornev2011} and its suitability in various fields of application like magnetometry~\cite{Schultze2003,Oppenlaender2003,Schultze2003a,Caputo2004,Caputo2005,Schultze2006,Polyakov2011} and rf electronics~\cite{Caputo2006,Caputo2007,Shadrin2007,Caputo2007a,Kalabukhov2008,Kornev2009a,Kornev2011}. {A problem with standard SQUID arrays, made of identical loop areas, is that although they are sensitive to changes in the magnetic flux density}, they do not work as absolute magnetometers~\cite{Haussler2001}. This renders this solution to the quantum sensing problem more difficult to integrate into real systems. {An improved solution to this problem exists and was put forward by Oppenländer~\textit{et al.}, who found that making the areas of the SQUID loops (in the arrays) incommensurate to each other~\footnote{No mutual division of these areas yield a rational number as a result.} leads to interference at each SQUID loop~\cite{oppenlanderNonEnsuremathPhi2000,oppenlanderTwoDimensionalSuperconducting2003}, resulting in an~\textit{anti-peak} response; a sharp dip in the voltage at zero flux, which allows the device to function as an absolute magnetometer.} This can be easily understood as the incommensurate distribution of areas causing destructive interference effects for all non-zero flux values~\cite{oppenlanderNonEnsuremathPhi2000,oppenlanderTwoDimensionalSuperconducting2003,Cybart2012,Coude2019}.

Indeed, the construction of SQIFs with a spread in the distribution of loop areas is certainly the cutting edge of the field~\cite{kornevSQIF2015}. However, an issue emerges when varying the sizes of the loops -- such as is done when spread in loop areas is introduced -- one also drastically changes the inductance $L$ of each individual SQUID loop. The central parameter this modifies is known as~$\beta_L$ and is given by~$\beta_L =  L I_C / \Phi_0$, where $I_C$ is the critical current of the component Josephson junctions, and~$\Phi_0$ is the {quantum} of magnetic flux~\cite{weinstockSQUIDSensorsFundamentals1996}. It has been determined experimentally that good control of~$\beta_L$ is critical to proper device performance~\cite{ClarkeBookC2}. Although there have been proposals to mitigate this, it still poses a significant technological challenge for the fabrication of large SQIFs with area spread. {Finding a solution to this problem will bring SQIFs one step closer to reaching the theoretically proposed quantum limit~\cite{Caves1982}}.

In this paper, we theoretically investigate (\textit{and experimentally confirm}) a possible new solution to the problem; the careful addition of `bare' superconducting loops -- loops of superconducting material containing \textbf{no} Josephson junctions -- {into a regular SQUID array to obtain a SQIF.} We do so by first defining the mathematical framework required to study {SQUID arrays} -- namely, the RSJ equations. By replacing some of the SQUID loops with bare superconducting loops, we then study how the conventional RSJ equations of motion are modified. Finally, we present some experimental results for {SQIF arrays obtained by inserting bare loops.}

\section{Model for SQUID arrays}
\label{sec::mod}

\begin{figure}[h]
    \centering
    \includegraphics[width=\linewidth]{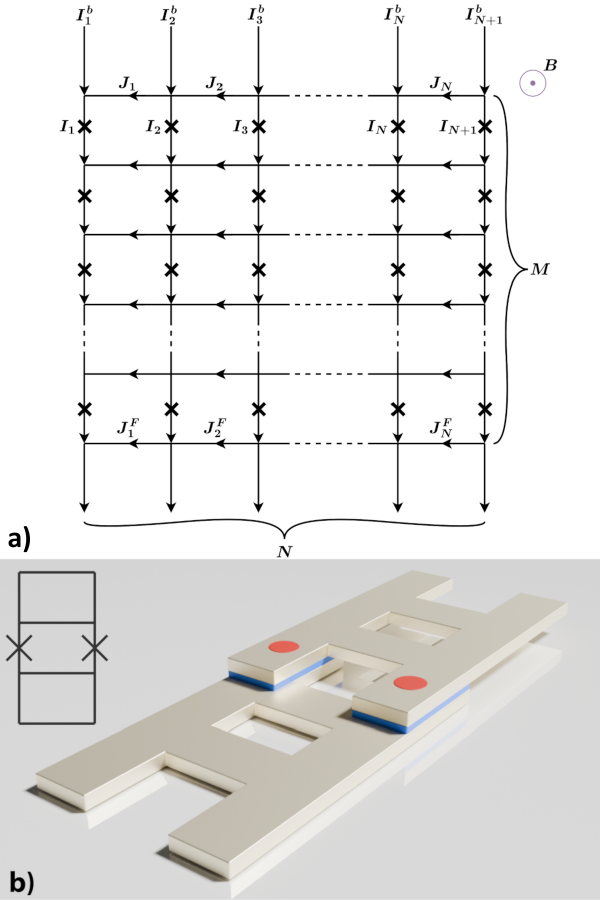}
    \caption{a)~Schematic of a prototypical SQIF containing both bare loops and SQUID loops. The numbering systems of the bias currents ($I^b_i$), the horizontal wires ($J_i$), the vertical wires ($I_i$), and the final row of horizontal wires ($I^F_i$) are labelled; the notation is identical to the circuit diagram of Ref.~\cite{galilabariasModelingTransferFunction2022} and is also described in more detail in Appendix~\ref{sec:2D}. In this notation, two vertical wires i-1 and i with Josephson junctions will surround the generic $i^{th}$ SQUID and the magnetic field $\vec{B}$ is directed coming off the page. Josephson junctions are represented by an 'X'. b)~Illustration of a unit-cell for the experimental array considered in this work. It consists of two bare loops surrounding a SQUID loop with the superconducting material shown in silver. Within the illustration, the red regions are vertical Josephson junctions, whilst the blue regions are made of insulating material. These junctions are normally shunted (not shown) in real circuits~\cite{oppenlanderTwoDimensionalSuperconducting2003,mitchell2DSQIFArrays2016,kornevSQIF2015}. A circuit schematic for this cell is also in the top left side. b) Is courtesy of A. Gardin.}
    \label{fig::Figure1}
\end{figure} 

As shown in~Fig.~\ref{fig::Figure1}.a), our model explicitly considers a {2D SQUID array}, with~$M$ rows, and~$N$ columns; hence, there {are} $N\cdot M$ total SQUIDs within the array, and $M\cdot(N+1)$ Josephson junctions. We assume the physical properties of each of these junctions, such as the resistance~$R$ and critical current~$I_C$ are constant across the array. The overall dynamics of the array is determined~\cite{oppenlanderTwoDimensionalSuperconducting2003, weinstockSQUIDSensorsFundamentals1996,galilabariasModelingTransferFunction2022} by taking into account the phase difference across each Josephson junction of the array,~$\underline{\varphi}$. Correspondingly, for the~$i$\textsuperscript{th} SQUID within the array, the relevant parameters are the phase differences across the two Josephson junctions, $(D\varphi)_i$, and the physical area of each SQUID loop $a_i$. {For simplicity, we consider the effects of self-inductances and nearest neighbour mutual inductances, ignoring further mutual inductance contributions~\cite{galilabariasModelingTransferFunction2022}.} \added{For simplicity, we ignore the effects of junction capacitance.}

\begin{figure}[h]
    \centering
    \includegraphics[width=\linewidth]{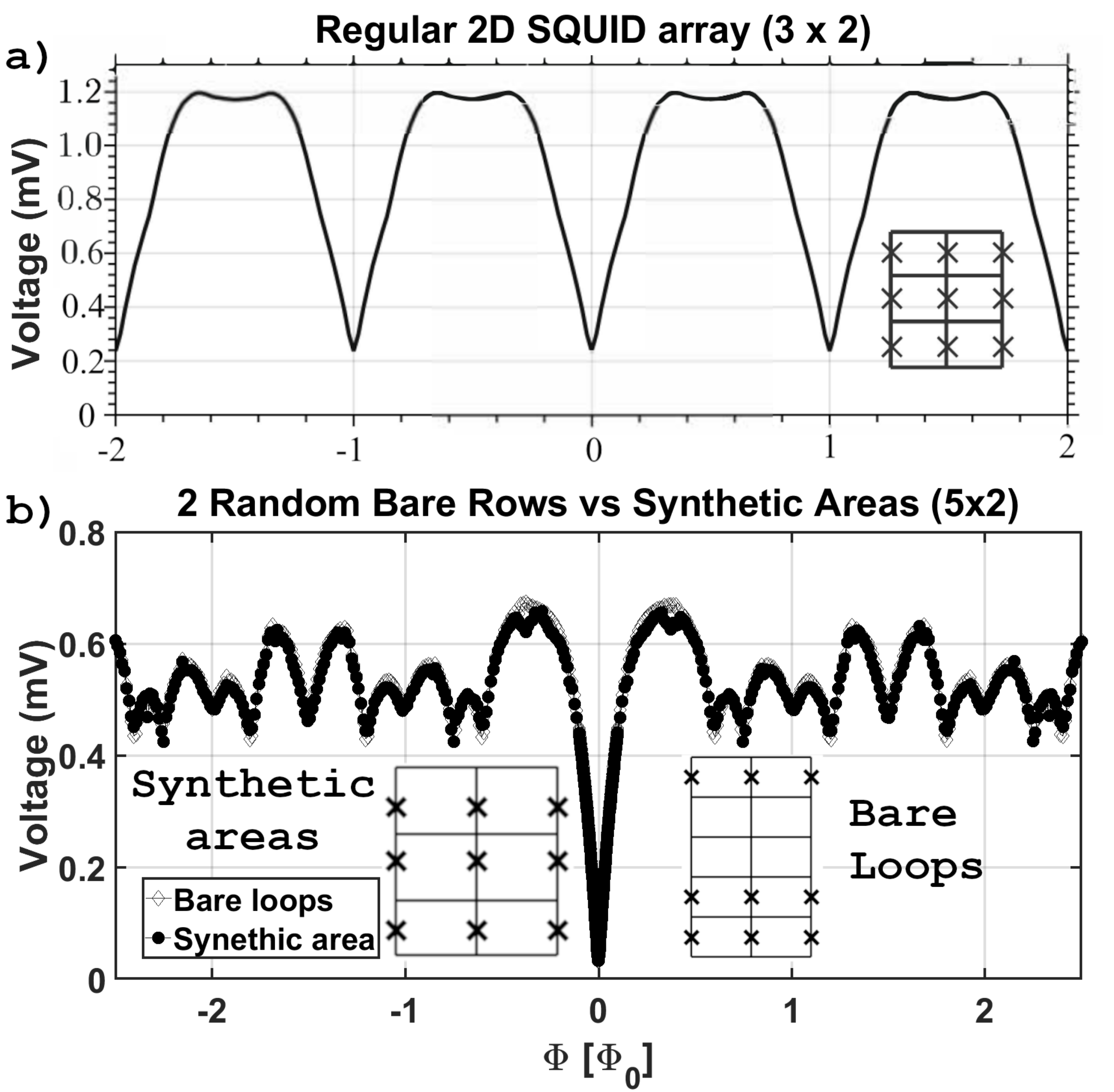}
    \caption{(a) The Voltage-Magnetic Flux (VMF) response of a $3\times 2$ array with equal loop areas is shown. (b) The VMF response of a $5\times 2$ array containing two rows of solely bare loops is shown with black diamonds; each loop has an equal area as depicted in the circuit diagram. The VMF response of a $3\times 2$ array where the loop area is given by the synthetic areas derived from the original circuit is overlain; this was performed by solving Eq.~\ref{eq::syntheticArea}. Both devices have the same inductances for each loop ({synthetic} or bare), shunt resistances of 9.6$\Omega$, {a device critical current of 250$\mu$A and an applied bias current of 270$\mu$A.}}
    \label{fig::SimData2}
\end{figure}     
\begin{figure}[h]
    \centering
    \includegraphics[width=\linewidth]{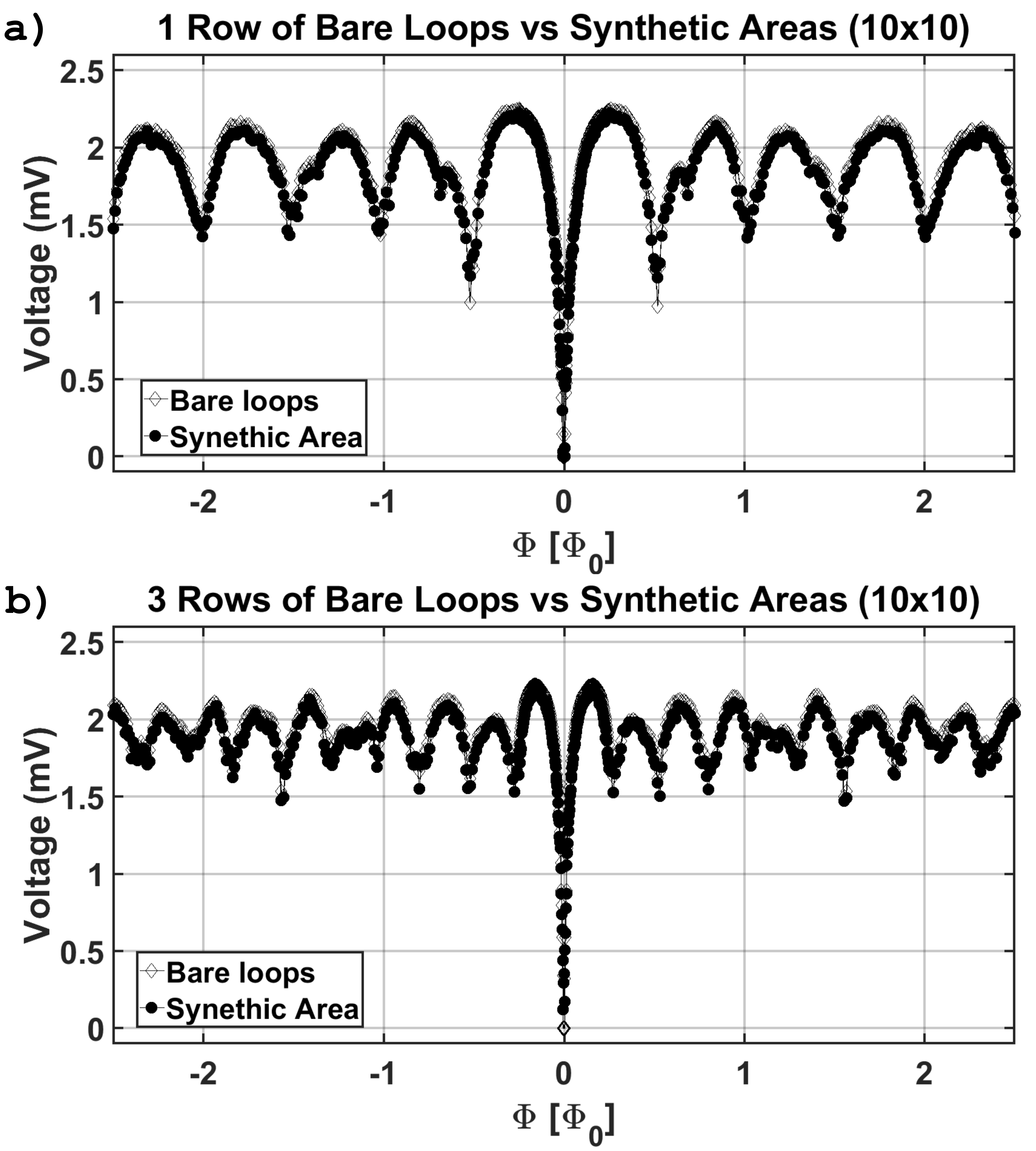}
    \caption{Shown with the black empty diamonds is the Voltage-Magnetic Flux (VMF) response of a $10\times 10$ array containing a) one or  b) three rows of solely bare loops, where each loop has an equal area. Shunt resistances of bare and synthetic devices in both a) and b) are 9.6$\Omega$ {and the devices have a critical current of $490\mu$A with data sets shown for a bias current of $510\mu$A.} Shown with the black full circles is the VMF response of arrays where the loop area is given by the synthetic areas derived from the two different configurations, a) and b); this was performed by solving Eq.~\ref{eq::syntheticArea}. For both these Figures, the differences between the bare loop and synthetic area approaches are minimal. Schematics of devices (a) and (b) may be found in Fig.~6 and Fig.~7 respectively of Appendix~\ref{sec::schematics}.}
    \label{fig::SimData3}
\end{figure}    

We begin by first providing a mathematical model to study {SQUID arrays by making use of the RSJ equations~\cite{tinkhamIntroductionSuperconductivity1975}, and will extend this to encapsulate a device containing bare loops as in Fig.~\ref{fig::Figure1}.b) in Section~\ref{sec::synthetic}}. Although we leave further details to Appendix~\ref{sec:2D} and references therein, the dynamics of the array can be summarised by solving~\cite{oppenlanderTwoDimensionalSuperconducting2003, weinstockSQUIDSensorsFundamentals1996,galilabariasModelingTransferFunction2022} the set of coupled equations as below;
\begin{equation}\label{eq::testRSJ}
    \frac{d\underline{\varphi}}{d\tau} = \underline{i}^b - \sin \underline{\varphi}  + Q \left[\underline{D\varphi}- \underline{a}\Phi\right]   \; ,
\end{equation} 
{where the underlines denote a vector, $Q$ is a matrix describing Kirchhoff's laws and inductances, $a$ gives the area of each SQUID loop {and $\Phi$ describes the flux through each loop, given in full by
\begin{equation}
    \Phi = 2\pi \frac{\phi}{\Phi_0},
\end{equation}
where $\phi$ is the magnetic flux density through a loop.} We have also defined a unit-less time variable}
\begin{equation}\label{eq::tau}
    \begin{aligned}
        \tau \equiv \frac{2\pi RI_Ct }{\Phi_0} \; .
    \end{aligned} 
\end{equation} 

Each quantity in Eq.~\ref{eq::testRSJ} is described in further detail within  Appendix~\ref{sec:2D}, however, each term can be ascribed to a physical origin~\cite{oppenlanderTwoDimensionalSuperconducting2003,weinstockSQUIDSensorsFundamentals1996,galilabariasModelingTransferFunction2022}: $\sin \underline{\varphi}$ is the current term due to the first Josephson relation, the\textbf{~$\partial_{\tau} \underline{\varphi}$}~=~$\frac{d\underline{\varphi}}{d\tau}$ is the derivative of the phase difference term due to the second Josephson relation, the~$\underline{i}^b$ term is the current through each Josephson junction due to the bias current being applied, and $Q \left[\underline{D\varphi}- \underline{a}\Phi\right] $ is the (total) current due to inductive contributions. Concerning the dimensions, the vectors $\underline{\varphi}$, and $\underline{i}^b$ contains $M\cdot(N+1)$ elements as this is the total number of Josephson junctions, whilst $\underline{D\varphi}$ and $\underline{a}$ contain $N\cdot M$ element as this is the total number of SQUIDs; $Q$, being a matrix, has a resulting dimension of $(M\cdot(N+1) \times N\cdot M)$.

Given the physical parameters of each Josephson junction, Eq.~\ref{eq::testRSJ} can be solved numerically to determine the phase difference across each junction. We can then substitute this solution back into Eq.~\ref{eq::testRSJ} to determine the voltage across each junction from the second Josephson relation
\begin{equation}\label{eq::JunctionVoltage}
    V_i(\tau) = RI_C \frac{d{\varphi_i}}{d\tau}  \; .
\end{equation} 
Finally, by averaging the voltage over parallel junctions
\begin{equation}\label{eq::par}
    \bar{V}(\tau) \equiv \frac{1}{N+1} \sum_i V_i(\tau)  \; ,
\end{equation} 
we can compute the experimentally measurable time-averaged voltage by
\begin{equation}\label{eq::TimeAveraged}
    \left< \bar{V} \right> \equiv\lim_{T\to \infty} \frac{1}{T}\int_0^T  \bar{V}(\tau')  \; d\tau' \; .
\end{equation}

\section{Synthetic area spread}\label{sec::synthetic}
In this section, we generalise the previously discussed model already present in the literature, such as the \added{works of} ref.~\cite{oppenlanderTwoDimensionalSuperconducting2003, galilabariasModelingTransferFunction2022}. Essentially, we will consider the dynamics of the array outlined in Sec.~\ref{sec::mod} after removing Josephson junctions such that there exist loops within the array containing no Josephson junctions; we will call such loops `bare' superconducting loops. We assume that of the previous $M\cdot(N+1)$ junctions, there are now $N_{\text{JJ}}$ Josephson junctions remaining within the array. Similarly, we denote the number of remaining `junction loops' -- namely, loops containing at least a single Josephson junction -- as $N_{\text{JL}}$, such that there are now $N\cdot M - N_{\text{JL}}$ bare loops within the array.

We now differentiate whether a quantity corresponds to either a bare loop, or to a junction loop, by adding a subscript $B$, or $J$, respectively. For example, we let $\underline{a}_J$ correspond to the vector containing the areas of all junction loops, whilst $\underline{a}_B$ corresponds to the area of all of the bare loops. The matrix $Q$ is then partitioned into the form
\begin{equation}\label{eq::Qpart}
    Q = \begin{bmatrix}
        Q_{JJ} & Q_{JB} \\ Q_{BJ}   & Q_{BB}                    
    \end{bmatrix}  \; ,
\end{equation} 
where the sub-matrix $Q_{JJ}$ has dimension $(N_{JJ} \times N_{\text{JL}})$ as it maps the flux through the junction loops into inductive current through the remaining Josephson junctions. We note that by definition both $\underline{\varphi}_{B} = 0$ and $\underline{D\varphi}_{B} = 0$ within this partition as there are no Josephson junctions to provide a {Josephson} phase difference. By partitioning each physical quantity in this manner, the equation of motion for the system is written as
\begin{equation}\label{eq::PartEq}
    \frac{d\underline{\varphi}_{J}}{d\tau} = \underline{i}^b_{J} - \sin \underline{\varphi_{J}}  + Q_{JJ} \left[\underline{D\varphi}_{J}- \underline{a}_{J}\Phi\right]  - Q_{JB} \underline{a}_{B}\Phi  
\end{equation} 
Although this equation is similar to Eq. \ref{eq::testRSJ}, the difference is the addition of the anomalous term $Q_{JB} \underline{a}_{B}\Phi$, which factors in the inductive current induced by the bare loops within the device.

We see from the structure of this equation that the matrix $Q_{JJ}$ maps the magnetic flux through the $N_{\text{JL}}$ junction loops into the current across the $N_{\text{JJ}}$ junctions; the precise nature of this mapping is constrained by Kirchhoff's law. However, as Kirchhoff's laws are overdetermined, then the rank of $Q_{\text{JJ}}$ is equal to $\min(N_{\text{JJ}}, N_{\text{JL}})$. For example, if there are two Josephson junctions shared between a single loop, then by Kirchhoff's law, the two currents are a function of the single flux value such that the rank of $Q_{JJ}$ is identically one. Similarly, if there are instead two Josephson junctions shared between, say, three superconducting loops, then by Kirchhoff's law, the current across the two junctions can be tuned independently by the extra flux value present such that $\text{rank}(Q_{JJ})=2$. The same logic applies to the total matrix $\left[Q_{JJ} \; Q_{JB}\right]$ such that 
\begin{equation}\label{eq::Rank}
    \text{rank}(\left[Q_{JJ} \; Q_{JB}\right]) = \text{rank}(Q_{JJ})  \; .
\end{equation}   
The purpose of illustrating this is that it implies that the columns of $Q_{JB}$ must be a linear combination of the columns of $Q_{JJ}$, otherwise~Eq. \ref{eq::Rank} would not be satisfied. The explicit restriction of $\text{rank}(Q)$ is shown in Appendix~\ref{sec::rank}. Hence, we can introduce a coefficient matrix $C$ such that
\begin{equation}\label{eq::CoeffMat}
    Q_{JB} = Q_{JJ}C  \; .
\end{equation} 

The matrix of coefficients may not be unique in that the matrix problem defined by Eq.~\ref{eq::CoeffMat} may be underdetermined depending on the precise location of where the Josephson junctions have been removed. Regardless, we can substitute this into Eq. \ref{eq::PartEq} to write the resulting equations of motion as
\begin{equation}\label{eq::FinalEq}
    \frac{d\underline{\varphi}_{J}}{d\tau} = \underline{i}^b_{J} - \sin \underline{\varphi_{J}}  + Q_{JJ} \left[\underline{D\varphi}_{J}- \left(\underline{a}_{J} + C\underline{a}_B\right)\Phi\right].
\end{equation} 
In Eq. \ref{eq::FinalEq} the area in the equations is no longer the physical area of each junction loop, given by~$\underline{a}_J$, but instead a synthetic set of areas, containing a contribution from the bare loops' areas~$\underline{a}_B$. We denote this synthetic set of areas by~$\underline{a}'$ 
\begin{equation}\label{eq::syntheticArea}
    \underline{a}' \equiv  \underline{a}_{J} + C\underline{a}_B  \; ,
\end{equation} 

Eq. ~\ref{eq::FinalEq} is a fundamental result of this work, as it demonstrates that the dynamics of the array is determined by the synthetic set of areas, rather than the physical junction loop areas. An immediate consequence of this is that if all of the physical areas are identical, one would expect a periodic VMF response from the array~\cite{oppenlanderNonEnsuremathPhi2000}. However, by simply removing Josephson junctions from the array, the relevant parameter becomes the synthetic set of areas, which may also be engineered as incommensurate. {While the inductances of the SQUID loops remain unchanged by this process, the concept of the $\beta_L$ parameter becomes less clear. It should be the subject of future work to explore if $\beta_L$ is a suitable parameter for such devices, or, if there are some other metrics by which performance should be measured. We do theorise, however, that as there is no need to physically modify the areas or inductances of each individual SQUID loop, the risk of degradation of the performances of these SQIFs should be reduced when their size increase to large N and M as discussed in Ref.~\cite{oppenlanderTwoDimensionalSuperconducting2003,galilabariasModelingTransferFunction2022}.}

To explore this result we study different devices. Without loss of generality, most of the results discussed in this work are for the case where only $I^b_{N/2+1}$ is $\neq$ 0, also called the central bias limit, and without any forms of noise included into the model. {The inductances used within the following theoretical simulations were obtained via commercial products such as \added{InductEx (version LTS v1.0.10)~\cite{Fourie2011}}, chosen to match the experimental parameters outlined in Appendix~\ref{sec::Methods}.} Initially, a small device consisting of six SQUID loops distributed in two columns and three rows is numerically considered. In Fig.~\ref{fig::SimData2}.a), we plot the time-averaged voltage for this array when every loop area is the same size. This is a well-known result; the response is characteristically periodic as {the SQUID loops interfere coherently}~\cite{oppenlanderTwoDimensionalSuperconducting2003}.

In Fig.~\ref{fig::SimData2} b), {we plot the response of two different devices: one, plotted with black diamonds, is a SQIF with five rows and two columns containing two rows of bare loops and all loops having the same area. Secondly, the data plotted with black full circles, are for a SQIF with three rows and two columns where loop areas are defined by Eq. 11. It is important to note that both the devices modelled in Fig.~\ref{fig::SimData2} b) have equal loop inductances for all loops.} For this example, by solving Eq.~\ref{eq::syntheticArea}, we find that the areas of the equivalent device are given by the following sequence; $\underline{a}' =\begin{bmatrix}
    \frac{3}{2} &  \frac{3}{2} &  \frac{3}{2} &  \frac{3}{2} & 1 & 1
\end{bmatrix}$, in units of the area used for the bare loops original structure. Fig.~\ref{fig::SimData2}.b) also displays {that due to the presence of the two rows of bare loops in the circuit, both responses have a width of the dip of around $0.5~\Phi_0$, which is half of the periodicity for the case without bare loops in Fig.~\ref{fig::SimData2} a).} This is a manifestation of the fact that, in our plots, we are using the magnetic flux through a single DC SQUID loop as a reference for the x-axis, even when extra bare loops are introduced into the circuit. This, in turn, means that each DC SQUID is synthetically varied by the presence of the bare loops. Hence, this section confirms that by the introduction of bare loops, a centrally peaked VFM response can be obtained, without the need to impose a physical spread in the areas.

In Fig.~\ref{fig::SimData3} another pair of devices where we take a $10\times 10$ SQIF and incorporate a) one or b) three rows of equal area bare loops is theoretically investigated. Again, solving Eq.~\ref{eq::syntheticArea}, we can find the sequence for the synthetic areas of the equivalent devices, which are shown in Appendix~\ref{sec:ArrayAreas}. Importantly, the devices modelled in Fig.~\ref{fig::SimData3} are just two examples of the many possible configurations that can be used to demonstrate the very powerful concept of synthetic area spread. The same devices as in Fig.~\ref{fig::SimData3} a) and b) are shown with a much broader Magnetic Flux range in Fig.~\ref{fig::SimData4}, found in Appendix~\ref{sec::Moreresults}. These extra simulations demonstrate that it is fairly easy to define a bare loop configuration, with no physical spread, but for which the equivalent synthetic spread results in a singular global minimum at zero flux. Hence, it is always possible to define devices that manifest an absolute zero, as opposed to that of a conventional  DC SQUID response~\cite{oppenlanderTwoDimensionalSuperconducting2003,mitchell2DSQIFArrays2016,kornevSQIF2015}, without the need of using performance degrading techniques. Lastly, the robustness of bare loop voltage vs synthetic area voltage against array size is also investigated in Fig.~\ref{fig::SimData5} of Appendix~\ref{sec::Moreresults}. It is shown that even for a SQIF containing $64\times 64$ DC SQUID loops and one, two or three rows of bare loop incorporated between each row of DC SQUIDs, the average Voltage Difference between the SQIF with bare loops and the corresponding synthetic area SQIF, calculated without any optimisation in the bias currents, is always maintained to a minimal value.

\section{Experimental Results}
\label{sec::Exp}

\begin{figure}[h]
    \centering
    \includegraphics[width=\linewidth]{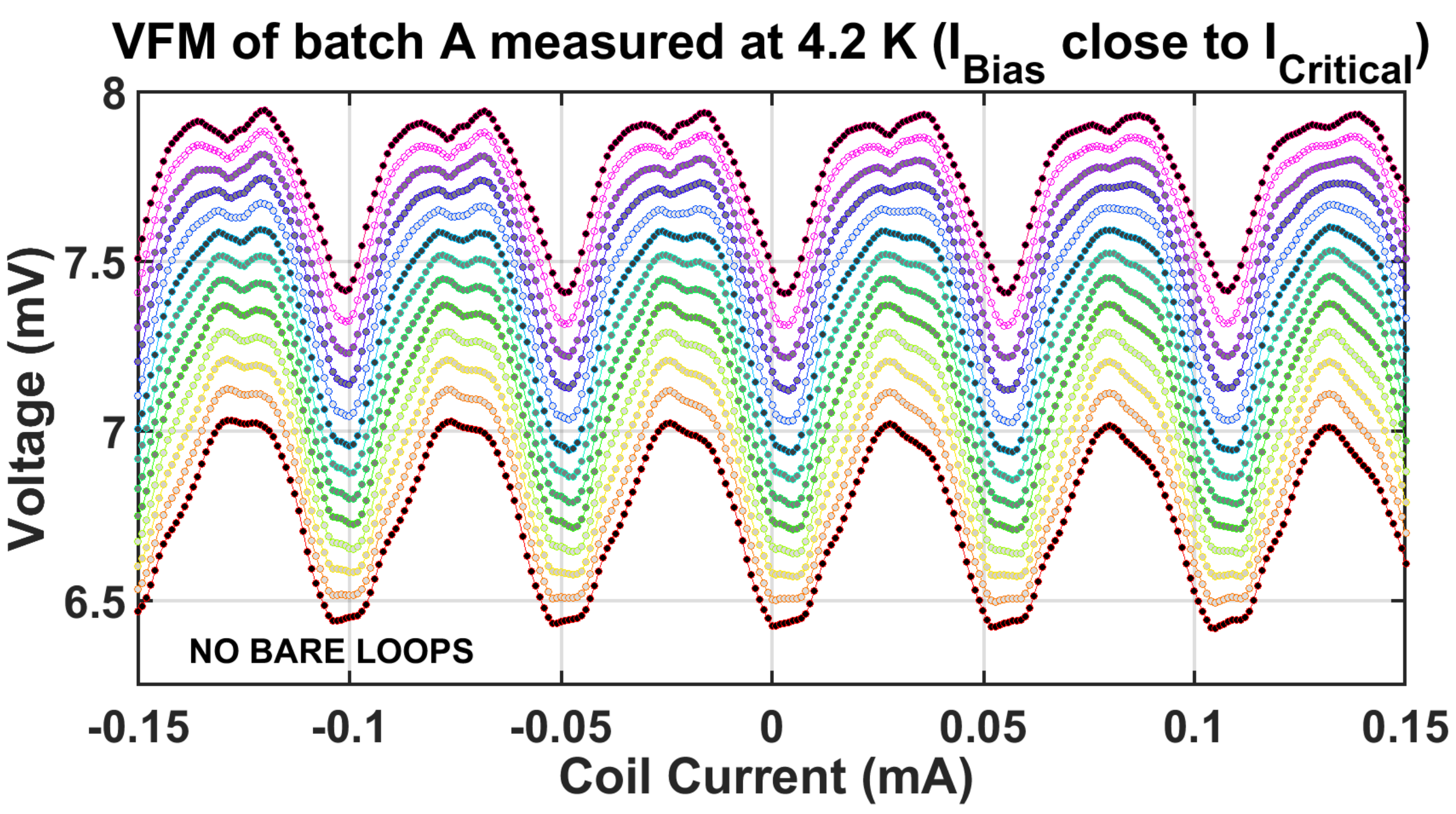}
    \caption{{Experimental VMF curves for different biases close to the critical one for a $16\times16$ {SQUID array}. {The device has a critical current of $835\mu$A, and the data sets shown are for bias currents in the range $838.5\mu A$ to $870\mu A$.} This device is from the same fabrication batch, referred to as batch A, as the one pictured in Fig.~\ref{fig::expData}. As expected, the VMF curves of the device are not anti-peaked.}}
    \label{fig::expDataB}
\end{figure} 
\begin{figure}[h] 
    \centering
    \includegraphics[width=\linewidth]{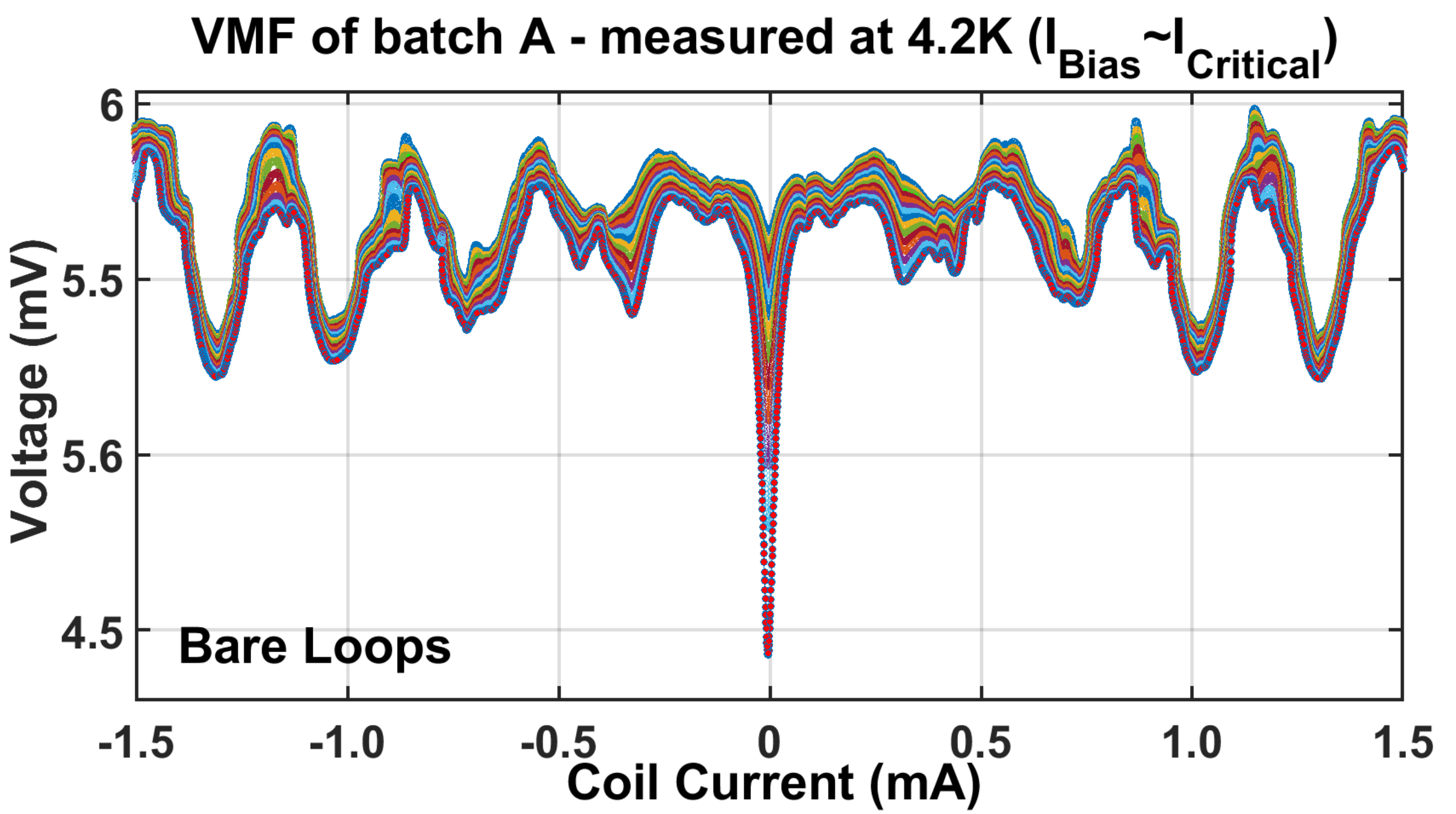}
    \caption{{Experimental VMF curves, for different biases close to the critical one. Results shown are from fabrication batch A and comprised of $46\times16$ SQIFs where sixteen rows contain Josephson junctions and two rows of bare loops between each row containing Josephson junctions. {Device has a critical current of approximately $820\mu$A and the data sets shown are for bias currents in the range $832\mu$A to $846\mu$A.} Although all the SQUID loop structures have identical areas, a strong anti-peak response visible indicates the presence of synthetic loop areas induced by the bare loops present in the circuit. The data for a device with an identical design but made in a different batch, batch B, demonstrated identical low temperature response and are shown in {Fig.~\ref{fig::expData2}} of Appendix~\ref{sec::Moreresults}.}}
    \label{fig::expData}
\end{figure} 

{Our theoretical predictions were experimentally explored in Fig.~\ref{fig::expDataB} and Fig.~\ref{fig::expData} where arrays without physical spread were fabricated. Both devices measured here were fabricated together as a part of batch A. Additional measurements taken from a separate fabrication batch, referred to as batch B, are found in Appendix~\ref{sec::Moreresults} to demonstrate reproducibility. The details of the fabrication methods, and respective device parameter values, for these devices are described in Appendix~\ref{sec::Methods}. Fig.~\ref{fig::expDataB} shows {a regular SQUID array} of sixteen rows and sixteen columns. The device shown in Fig.~\ref{fig::expData} is made by the same fabrication process, but while still containing sixteen columns and sixteen rows with Josephson junctions, two rows without junctions are placed between each with junctions. This results in an array of size $46\times16$, where loops containing Josephson junctions are twice the size of those without.} 


To probe the voltage response of these devices, an external superconducting coil was used to apply a homogeneous magnetic field through each of the arrays. {A conventional cryogenic 4-point configuration has been used to apply a bias current in order to bring the device towards the correct resistive mode for device operation, but also to allow the measurement of the voltage drop across the array.} It is worth noting that the coil current range of the x-axis for every experimental plot is limited by the maximum amount of current that can be sustained by the coils used to generate the applied magnetic fields. It is also important to outline here that the experimental setup is similar to the theoretical assumption of the central bias limit discussed in Fig.~\ref{fig::SimData2} and Fig.~\ref{fig::SimData3}. By sweeping the external current through the coil, the characteristic Voltage-Magnetic Flux (VMF) curve for each device can be measured. The typical experimental VMF response for our devices is plotted in Fig.~\ref{fig::expData}. We note that (as also shown in Appendix~\ref{sec::Moreresults}) this curve was highly reproducible with multiple devices with identical design chosen at random across the wafer. For the curves shown, the bias current was tuned such that the best anti-peak response was observed. The striking behaviour of this measurement is the appearance of a significant anti-peak. {It is also noted that due to the device of Fig.~\ref{fig::expData} experiencing a larger effective loop area, more current must be driven through the coil current to experience any periodic nature.} Indeed, there are no additional significant minima as the strength of the external magnetic field is increased. Concerning the discussion in Sec.~\ref{sec::synthetic}, Fig.~\ref{fig::expDataB} (and identically in Fig.~\ref{fig::expData2} found in Appendix~\ref{sec::Moreresults}), we consider a separate {SQUID array} which contains no bare loops and observe that no central anti-peak is observed in this case. Hence, we have illustrated both theoretically and experimentally that the addition of bare loops can be used to dramatically modify the VMF response in our {SQUID arrays}. In this section, the experimental data represent a qualitative realisation of the effect, as we \added{do not} exclude that other effects (e.g. device imperfections) may have played a minor role in the way experimental results have manifested. However, these results confirm that the physics that dominates the experimental responses is linked to the bare loops and the corresponding synthetic areas generated.

\section{Summary and Outlook}
\label{sec::Con}

This work demonstrates a novel method that can be {used to modify 2D SQUID arrays such that they operate as SQIFs, showing the characteristic anti-peak response,} without degrading their performance. It not only has the merit to broaden the state-of-the-art theoretical results, but most importantly, provides a clear analytical and numerical explanation on how to reproduce this central effect, opening the way to its routine use. The key message of this work is that it provides a fully analytical, numerical and experimental demonstration of the new concept of synthetic area spread. By distributing bare loops of superconducting materials and incorporating them ad hoc in the superconducting arrays of DC SQUIDs, these systems {are turned into SQIFs, suitable to be used as absolute magnetometers} without degrading \added{individual SQUID cell performance}; hence, this may finally open the way to ultra-performant integrated sensors for electromagnetic signals. In conclusion, the importance of this work is that it has the potential to finally unlock the ability of SQIF technology towards the quantum limit of detection of electromagnetic signals.

\section{Data availability}
\label{sec::Data}
All relevant data presented in the paper are available upon reasonable request.

\section{Code availability}
\label{sec::Code}
All computer code used in the paper is available upon reasonable request.

\section*{Acknowledgments}
\label{sec::Ack}
The research was partially supported by the University of Adelaide via the CAS funding scheme. R. D.M., J. L. M. and G. C. T are supported by the Australian Commonwealth Government through Research Training Program Scholarships. J. L. M. is also funded via the Ian Croser Research Scholarship in Engineering at the University of Adelaide. Giuseppe C. Tettamanzi acknowledge funding from the Australia’s Economic Accelerator Innovate Program scheme project number IV240100119. G.C.T. conceived the project. R. D.M. performed the analytical and numerical calculations with input from all authors. The concept of Synthetic Spread of the Areas and the subsequent idea of Mixed SQUID arrays are the subject of an international patent application no. PCT/AU2025/050991. We acknowledge discussions with T. Whittaker concerning the nature of the bare loops $\&$ the interpretation of the exp. data and A. Gardin for the production of Fig. \ref{fig::Figure1}b).

\bibliography{references.bib}

\appendix

\section{\label{sec:2D}Mathematical model of SQIFs}

To determine the equivalence between SQIFs containing bare loops with no area spread, and those with area spread and no bare loops, we must first choose a formalism to work with. To this end we will make use of a circuit theory approach utilising the RSJ equations \cite{tinkhamIntroductionSuperconductivity1975}. We will adopt the matrix formalism introduced in  Ref.~\cite{galilabariasModelingTransferFunction2022}, however, we will extend their results to factor in the introduction of bare superconducting loops.

In Fig.~\ref{fig::Figure1} a) we depict a schematic of a {regular 2D SQUID array}. It consists of $M$ superconducting loops in the vertical axis, and $N$ loops in the horizontal axis. The modifications to be made to this schematic is that while some of these loops of superconducting wire have two Josephson junctions along their vertical sides, hence forming SQUID loops; the remaining loops do not contain Josephson junctions, we denote these as 'bare' superconducting loops. Indeed, this generalised mixed {SQUID array} can be considered as a standard {SQUID array} \added{with bare superconducting loops inserted at equal intervals}. In the specific example shown in Fig.~\ref{fig::SimData2} b), we see that the first, fourth, and final rows consist of SQUID loops, whilst the second and third rows contain solely bare loops. This configuration in Fig.~\ref{fig::SimData2} b) is one example and other more specific cases described in this papers. It is indeed important to outline here that our modelling approach is quite general and that the examples discussed in this work will represent only a sub-set of what is possible to be modelled.

We denote the current through the $k$\textsuperscript{th} vertical wire segments as $I_k$, where, reading from left to right and top to bottom, we number each wire sequentially; only the first $N+1$ vertical wire segments are labelled in Fig.~\ref{fig::Figure1}. For the horizontal wire segments we denote them by $J_k$ and also label them sequentially, however, we distinguish the final row of horizontal currents by labelling them as ${J^F}_k$. We will denote the vectors containing each of these elements by the underlined quantities $\underline{I}$, $\underline{J}$, $\underline{{I^b}}$, $\underline{{J^F}}$. Concerning dimensions: $\underline{I}$ contains $M\cdot(N+1)$ elements, $\underline{J}$ contains $N\cdot(M-1)$ elements, $\underline{{J^F}}$ contains $N$ elements, and we pad $\underline{{I^b}}$ with $(N+1)\cdot(M-1)$ zeros ($\underline{{I^b}} \equiv [{I^b}_1,\ldots,{I^b}_{N+1},0,\ldots,0]^t$) such that the resulting vector has the same length as $\underline{I}$.

Within a circuit theory, one can construct Kirchhoff's current conservation laws at each vertex of the circuit. For example, the conservation condition for the vertex located at the intersection between the $i$\textsuperscript{th} column and the  $j$\textsuperscript{th} row, assuming it is not lying on the edge of the graph, is given by 
\begin{equation}\label{eq::Kirchhoff}
    I_{i+(j-1)(N+1)} = J_{i+(j-1)N} + I_{i+(j-2)(N+1)} - J_{i-1+(j-1)N} \; .
\end{equation} 
Following the work of Ref. \cite{galilabariasModelingTransferFunction2022}, we introduce coefficient matrices ${K^I}$ and ${K^J}$ consisting entirely of $-1$, $0$, and $1$ such that the Kirchhoff conservation condition for every node can be represented as the matrix equation
\begin{equation}\label{eq::Kirchhoff matrix}
    {K^I}\underline{I} = {K^J}\underline{J} + \underline{{I^b}}  \; .
\end{equation} 
where $K^I$ has dimensions $[M\cdot(N+1) \times M\cdot(N+1)]$ and $K^J$ has dimensions $[M\cdot(N+1) \times N\cdot(M-1)]$.

As we distinguished the final row of horizontal currents, we must also introduce further Kirchhoff matrices $N_I$, with dimension $[N\times M\cdot(N+1)]$, and $N^F$, with dimension $[N \times N]$, defined such that the current conservation laws for the final row of vertices are contained within the system of equations
\begin{equation}\label{eq::FinalRowKirchhoff}
    \underline{J^F} = N^I\underline{I} + N^F\underline{I^F}   \; .
\end{equation} 
where $\underline{I^F} \equiv \left[{I^b}_1, \ldots, {I^b}_{N}\right]^t$ bundles the currents leaving the device into a vector.

The current through each wire segment will in-turn produce an inductive flux contribution through each superconducting loop, which we denote by the vector $\underline{\Phi}^L$ containing $N\cdot M$ elements. To determine the flux contributions resulting from these vertical ($\underline{I}$), horizontal ($\underline{J}$), and final row ($\underline{J}^F$) we must introduce inductance matrices $L^I$, $L^J$, $L^F$ respectively. Although the precise form of these matrices will be determined by the physical model for the inductances considered i.e. kinetic inductances, geometric inductances, they give the resulting inductive contribution to the flux through each loop by
\begin{equation}\label{eq::InductiveFlux}
    \underline{\Phi}^L = L^I \underline{I} +L^J \underline{J}+L^F \underline{J}^F\; .
\end{equation} 
Due to the {flux} quantisation condition for each superconducting loop, this inductive contribution to the flux is balanced by both the external flux through that loop, $\Phi^{\text{ext}}$, and any Josephson junction phase differences $\varphi$ present within that loop. However, in a bare superconducting loop there are no Josephson junction phases to balance the external flux; this is the critical distinction between a bare loop and a SQUID loop. Without the loss of generality, for the $i$\textsuperscript{th} loop in the array surrounded generically by the $k-1$ and $k$ vertical wire segments, the corresponding {flux} quantisation conditions through the SQUID loop and the bare loops are given by
\begin{equation}\label{eq::PhaseQuantisation}
    \begin{aligned}
        \text{SQUID :} &\quad \Phi^{\text{ext}} + \Phi^L_{i} = (D\varphi)_i\; +2\pi n_i , \\
        \text{bare :} &\quad \Phi^{\text{ext}}+ \Phi^L_{i} = 2\pi m_i \; .
    \end{aligned} 
\end{equation} 
where $(D\varphi)_i\equiv \varphi_{k}-\varphi_{k-1}$, {we have used the (integer) quantum numbers $n_i$ and $m_i$ and $\Phi^{\text{ext}/L}$ are redefined as
\begin{equation}
    \Phi^{\text{ext}/L}=2\pi\frac{\phi^{\text{ext}/L}}{\Phi_0}.
\end{equation}
In order to simplify our model we merge the $n_i$ into $(D\varphi)_i$, which are defined modulo $2\pi$, and drop the $m_i$.}

At the expense of introducing more notation, it is useful to introduce sets $\mathcal{S}$ and $\mathcal{J}$ defined as
\begin{equation}\label{eq::set}
    \begin{aligned}
        \mathcal{S} &\equiv \left\{ i \,|\, \text{the } i\textsuperscript{th} \text{ loop is a SQUID} \right\}  \; ,  \\
        \mathcal{J} &\equiv \left\{ k \,|\, \text{the } k\textsuperscript{th} \text{ vertical wire forms part of a SQUID} \right\}  \; ,
    \end{aligned} 
\end{equation} 
to allow us to distinguish notationally between wires and loops forming SQUID loops. The number of elements of these sets are precisely the number of SQUID loops, and the number of Josephson junctions, which we label as $N_{\text{SQUID}}$ and $N_{\text{JJ}}$ respectively. We can now compactly write the {flux} quantisation conditions for SQUIDS and bare loops defined by Eq. \ref{eq::PhaseQuantisation} in vector form as
\begin{equation}\label{eq::PhaseQuantVectir}
    \underline{\Phi}^{\text{ext}} + \underline{\Phi}^L = \underline{\theta} \; ,
\end{equation} 
where we have introduced a generalised difference vector $\underline{\theta}$ whose elements are defined by
\begin{equation}\label{eq::theta}
    \theta_i = 
    \begin{cases}
        (D\varphi)_i   & i \in \mathcal{S}  \; , \\
        0   & i \notin \mathcal{S}  \; .
    \end{cases} 
    \end{equation} 
With the aim of rewriting the inductive contribution within Eq. \ref{eq::PhaseQuantVectir} as a function of the currents through the device, we combine the Kirchhoff matrices defined previously with the inductive flux contribution defined in Eq. \ref{eq::InductiveFlux} to express the {flux} quantisation condition as a function of solely the horizontal currents $\underline{J}$
\begin{equation}\label{eq::KirchoccPhase}
    \underline{\Phi} + L\underline{J} = \underline{\theta}  \; ,
\end{equation} 
where 
\begin{equation}\label{eq::generalMatrices}
    \begin{aligned}
        L &\equiv L^J + (L^I + L^FN^I){K^I}^{-1}K^J    \; ,\\
        \underline{\Phi} &\equiv  \underline{\Phi}^{\text{ext}} +  (L^I + L^FN^I){K^I}^{-1}\underline{I}^b + L^FN^F \underline{I}^F \; . \\
    \end{aligned} 
\end{equation} 
The dynamics of each Josephson junction is a function of not only both the phase difference across that junction, $\varphi_k$, and the current $I_k$, but also the physical parameters such as the resistance $R$, and the critical current $I_C$ of each junction. Assuming identical over-damped Josephson junctions, the relationship between the current through each junction and the phase difference $\varphi$ is given by the RSJ model \cite{tinkhamIntroductionSuperconductivity1975}
\begin{equation}\label{eq::RSJ}
    \frac{I_k(t)}{I_C} = \sin\varphi_k(t) + \frac{\Phi_0}{2\pi RI_C} \frac{d\varphi_k}{dt}  \; , \; k\in \mathcal{J} \; .
\end{equation} 
Note that we have omitted a noise and further capacitative term, which, although are standard terms to include for the purposes of modelling, are irrelevant in proving the equivalence between arrays containing bare loops and no spread, and arrays containing area spread but no bare loops.

As the RSJ equations require only the currents through the Josephson junctions, rather than through every wire segment, we introduce a mask matrix ${P}$ with dimensions $[N_{\text{JJ}} \times  M\cdot(N+1)]$ which can project vectors defined over every wire segment onto vectors defined only over wire segments containing a Josephson junction. The elements of this projection matrix are defined as
\begin{equation}\label{eq::Mask}
    {P}_{ij} = \delta_{j,\mathcal{J}_i}  \; ,
\end{equation} 
such that the vector of currents through each Josephson junction is given by {$P\underline{I}$}. Substituting in Eq. \ref{eq::KirchoccPhase} and the Kirchhoff matrices, taking account of the presence of the bias currents $\underline{I}^b$, and through the use of the projection matrix {$P$}, we can now express the RSJ equations defined in Eq. \ref{eq::RSJ} as        
\begin{equation}\label{eq::RSJmodified}
    \frac{d\underline{\varphi}}{d\tau} = \underline{C} + KL^{-1} \left[\underline{\theta}- \underline{\Phi}\right]  - \sin \underline{\varphi}  \; ,
\end{equation} 
where we have defined
\begin{equation}\label{eq::FinalDefn}
    \begin{aligned}
        \tau     &\equiv \frac{2\pi RI_Ct }{\Phi_0} \; ,\\
        \underline{C}      &\equiv  ({P}{K^I}^{-1}{\underline{I}^b}) / I_C  \; ,\\
        K   &\equiv ({P}{K^I}^{-1}{K^J}) / I_C \; .
    \end{aligned} 
\end{equation} 
to simplify the resulting equation. Eq. \ref{eq::RSJmodified} is the equation of motion for a 2D array containing bare loops, whereupon all dynamics can be evaluated by solving this equation.

\section{\label{sec::schematics}Device schematics}

{In this section we provide a visualisation of the device schematics used to produce Fig.~\ref{fig::SimData3}.}

\begin{figure}[h]
    \centering
    \includegraphics[width=\linewidth]{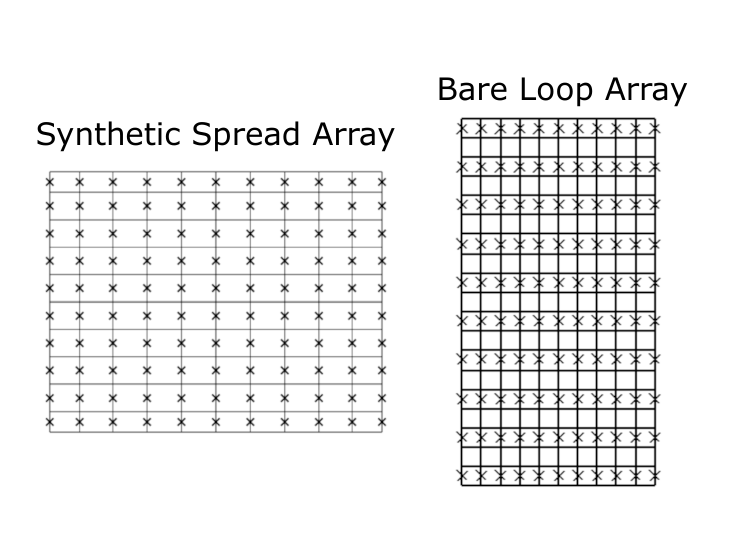}
    \label{fig::fig3aSchem}
    \caption{{Pictured here is the schematics of the two devices pictured in Fig.~\ref{fig::SimData3} a). Josephson junctions are marked by an X and the area of loops in the synthetic spread array are to scale relative to one another as found by Eq.~\ref{eq::syntheticArea}.}}
\end{figure}

\begin{figure}[h]
    \centering
    \includegraphics[width=\linewidth]{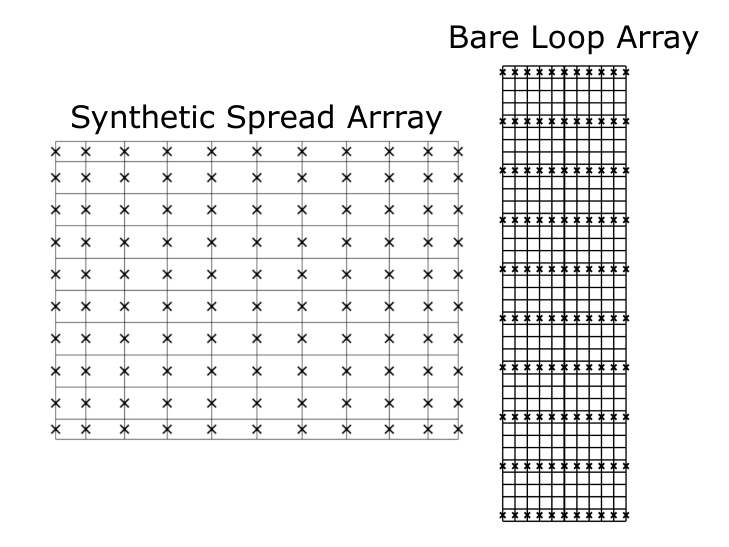}
    \label{fig::fig3bSchem}
    \caption{{Pictured here is the schematics of the two devices pictured in Fig.~\ref{fig::SimData3} b). Josephson junctions are marked by an X and the area of loops in the synthetic spread array are to scale relative to one another as found by Eq.~\ref{eq::syntheticArea}.}}
\end{figure}

\section{\label{sec::rank}Rank of sub-matrices}

In this section we consider the rank of the sub-matrices of $Q$ defined by
\begin{equation}\label{eq::defQ}
    Q \equiv {K^I}^{-1} K^J L^{-1}  \; ,
\end{equation} 

Importantly, $K_I$ and $L$ are invertible and hence full rank. As a result the rank of $Q$ is simply given by then rank of $K_J$
\begin{equation}\label{eq::rankEquiv}
    \text{rank}(Q) = \text{rank}(K_J)  \; ,
\end{equation} 

\section{\label{sec::Methods}Fabrication Methods}

The experimental data shown is of several device fabricated by SEEQC Ref.~\cite{SEEQC2024} using a typical fabrication process for niobium-based superconducting integrated circuits. This uses only refractory materials, with the exception for a Pd/Au metallisation layer used for contact pads. Niobium is used as the superconducting material due to its comparably high critical temperature, electrical and thermal stability, and ability to be thermally cycled many times without degradation. Niobium/Aluminum-Oxide/Niobium Josephson tunnel junctions are made by depositing an in-situ trilayer across the entire wafer and subsequently defining junction areas by deep-UV photolithography and etching. This method yields good uniformity and reproducibility of junction parameters. The critical current density of Nb/AlOx/Nb tri-layer associated with these devices is 4.5 kA/cm2. The sheet resistance of the Ti/PdAu/Ti resistive layer (R2) is 4.0 ohms/sq. The wafer used for the devices is fabricated on a 150-mm diameter (6-inch) high resistivity Si wafers.

Many different devices, starting from single Josephson Junction to DC SQUIDs have been fabricated and measured, and have demonstrated very high reproducibility as expected for this technology~\cite{SEEQC2024}. These devices were fabricated with identical array structure, SQUID loop area and bare loop area, but two distinct junction topologies; The first, referred to as batch A, of these have a shunt resistance of 9.6$\Omega$ and single junction critical currents of 55{$\mu$}A. The second, referred to as batch B, has a shunt resistance of 7.2$\Omega$ and single junction critical currents of 70{$\mu$}A. Numerous devices were fabricated with inductances varying between a few pH to 10pH. \added{All Josephson junction also have an internal capacitance in the range of $50$-$100$fF, and have a junction area of $2\mu$m$^2$.} A total of around ten devices with no spread have been measured at temperatures between 4.2 K to 8 K; two of them seem to be not responding to both Voltage Current and Voltage Flux measurement when studied.

The data of most all the others devices demonstrates high reproducibility and are discussed either in the main document or in these Appendices. Hence, approximatively 80 $\%$ of the devices used for this study are operating correctly (while the others are simply not operating at all). Conventional ultra-low noise setup is used to measure Voltage-Current and Voltage-Magnetic Flux curves. Many devices of different geometries have been successfully measured, although in this work we are focusing on the ones that have used the bare loop configuration.

\section{\label{sec:ArrayAreas}Synthetic areas}

The sequences for the synthetic area spreads for Fig.~\ref{fig::SimData3} are listed below. For the device with 1 row of bare loops pictured in Fig.~\ref{fig::SimData3} a) the sequence of synthetic areas is given by
$$
    \begin{bmatrix}
        1.36 & 1.46 & 1.49 & 1.49 & 1.49 & 1.49 & 1.49 & 1.49 & 1.46 & 1.36 \\ 
        1.73 & 1.92 & 1.98 & 1.99 & 1.99 & 1.99 & 1.99 & 1.98 & 1.92 & 1.73 \\ 
        1.73 & 1.92 & 1.98 & 1.99 & 1.99 & 1.99 & 1.99 & 1.98 & 1.92 & 1.73 \\ 
        1.73 & 1.92 & 1.98 & 1.99 & 1.99 & 1.99 & 1.99 & 1.98 & 1.92 & 1.73 \\ 
        1.73 & 1.92 & 1.98 & 1.99 & 1.99 & 1.99 & 1.99 & 1.98 & 1.92 & 1.73 \\ 
        1.73 & 1.92 & 1.98 & 1.99 & 1.99 & 1.99 & 1.99 & 1.98 & 1.92 & 1.73 \\ 
        1.73 & 1.92 & 1.98 & 1.99 & 1.99 & 1.99 & 1.99 & 1.98 & 1.92 & 1.73 \\ 
        1.73 & 1.92 & 1.98 & 1.99 & 1.99 & 1.99 & 1.99 & 1.98 & 1.92 & 1.73 \\ 
        1.73 & 1.92 & 1.98 & 1.99 & 1.99 & 1.99 & 1.99 & 1.98 & 1.92 & 1.73 \\ 
        1.36 & 1.46 & 1.49 & 1.49 & 1.49 & 1.49 & 1.49 & 1.49 & 1.46 & 1.36
    \end{bmatrix}
$$
Similarly for the device with 3 rows of bare loops pictured in Fig.~\ref{fig::SimData3} b), the sequence of synthetic areas is given by 
$$
    \begin{bmatrix} 
        1.80 & 2.17 & 2.34 & 2.41 & 2.44 & 2.44 & 2.41 & 2.34 & 2.17 & 1.80 \\ 
        2.60 & 3.34 & 3.68 & 3.83 & 3.89 & 3.89 & 3.83 & 3.68 & 3.34 & 2.60 \\
        2.60 & 3.34 & 3.68 & 3.83 & 3.89 & 3.89 & 3.83 & 3.68 & 3.34 & 2.60 \\
        2.60 & 3.34 & 3.68 & 3.83 & 3.89 & 3.89 & 3.83 & 3.68 & 3.34 & 2.60 \\
        2.60 & 3.34 & 3.68 & 3.83 & 3.89 & 3.89 & 3.83 & 3.68 & 3.34 & 2.60 \\
        2.60 & 3.34 & 3.68 & 3.83 & 3.89 & 3.89 & 3.83 & 3.68 & 3.34 & 2.60 \\
        2.60 & 3.34 & 3.68 & 3.83 & 3.89 & 3.89 & 3.83 & 3.68 & 3.34 & 2.60 \\
        2.60 & 3.34 & 3.68 & 3.83 & 3.89 & 3.89 & 3.83 & 3.68 & 3.34 & 2.60 \\
        2.60 & 3.34 & 3.68 & 3.83 & 3.89 & 3.89 & 3.83 & 3.68 & 3.34 & 2.60 \\
        1.80 & 2.17 & 2.34 & 2.41 & 2.44 & 2.44 & 2.41 & 2.34 & 2.17 & 1.80  
    \end{bmatrix}
$$

\section{\label{sec::Moreresults}More Results}

\begin{figure}[h]
    \centering
    \includegraphics[width=\linewidth]{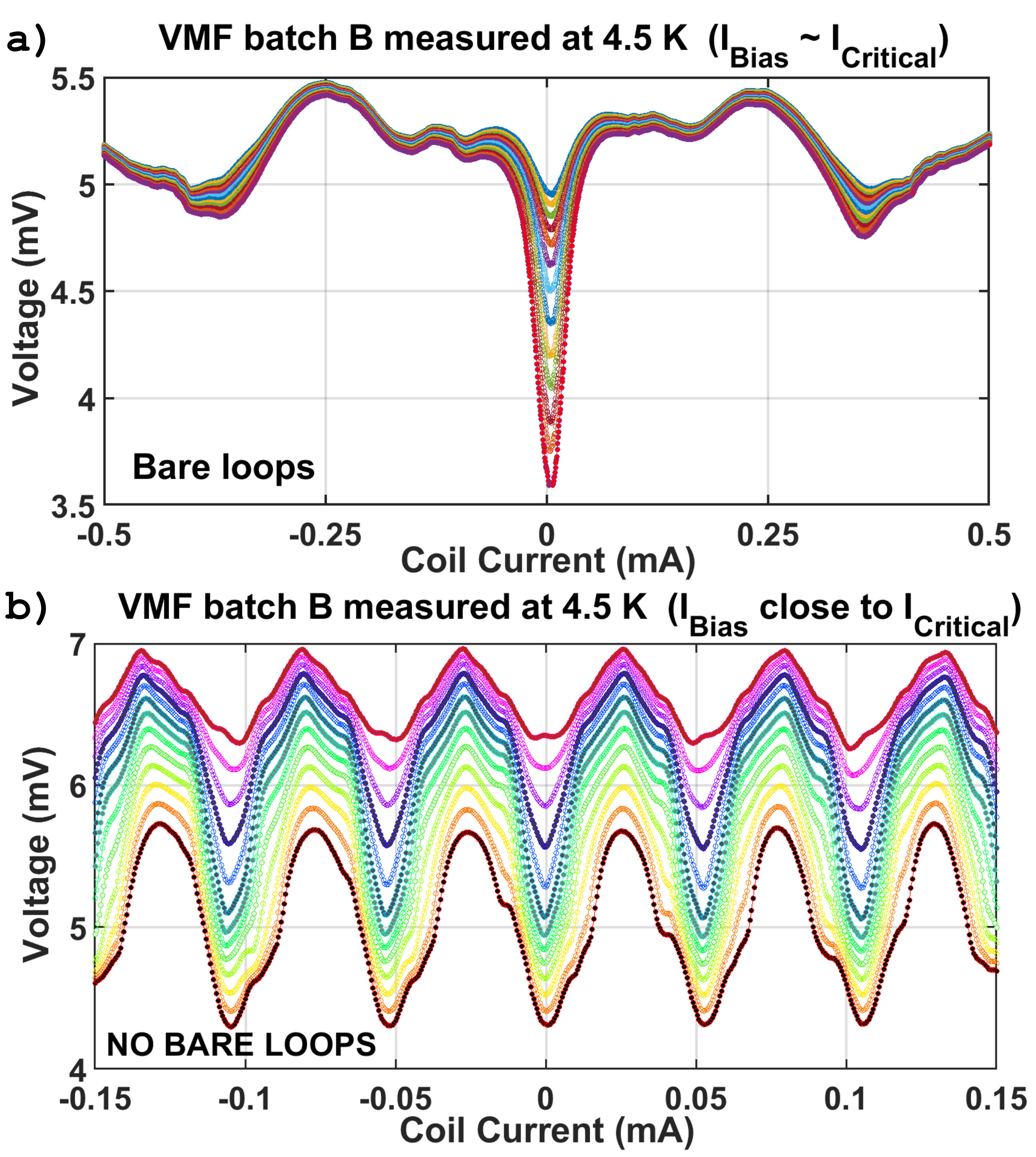}
    \caption{These experimental results show additional measurements for batch B. Batch B contains devices of identical design as in in Fig.~\ref{fig::expDataB} and Fig.~\ref{fig::expData} with different junction parameters as described in Appendix~\ref{sec::Methods}. (a) Shows a $46\times16$ SQIF for which there are two bare rows between each row with Josephson junctions. {The device had a device critical current of approximately $1.01$mA and bias currents are in the range of $1.075$mA to $1.083$mA}. (b) Shows a $16\times16$ {SQUID array} of the same parameters as (a) with no bare rows, {the device had a critical current of approximately $800\mu$A and bias currents in the range of $1.07$mA to $1.1$mA}. Contrary to (a), but as expected, there is no visible anti-peak.}
    \label{fig::expData2}
\end{figure}

{More theoretical and experimental results linked to the discussion in the main section of the paper are listed below. The experimental results of Fig.~\ref{fig::expData2} show additional measurements taken from similar devices as in Fig.~\ref{fig::expDataB} and Fig.~\ref{fig::expData} but fabricated with different junction values in what is referred to as batch B.}

%

%
%

%
\begin{figure}[h]
    \centering
    \includegraphics[width=\linewidth]{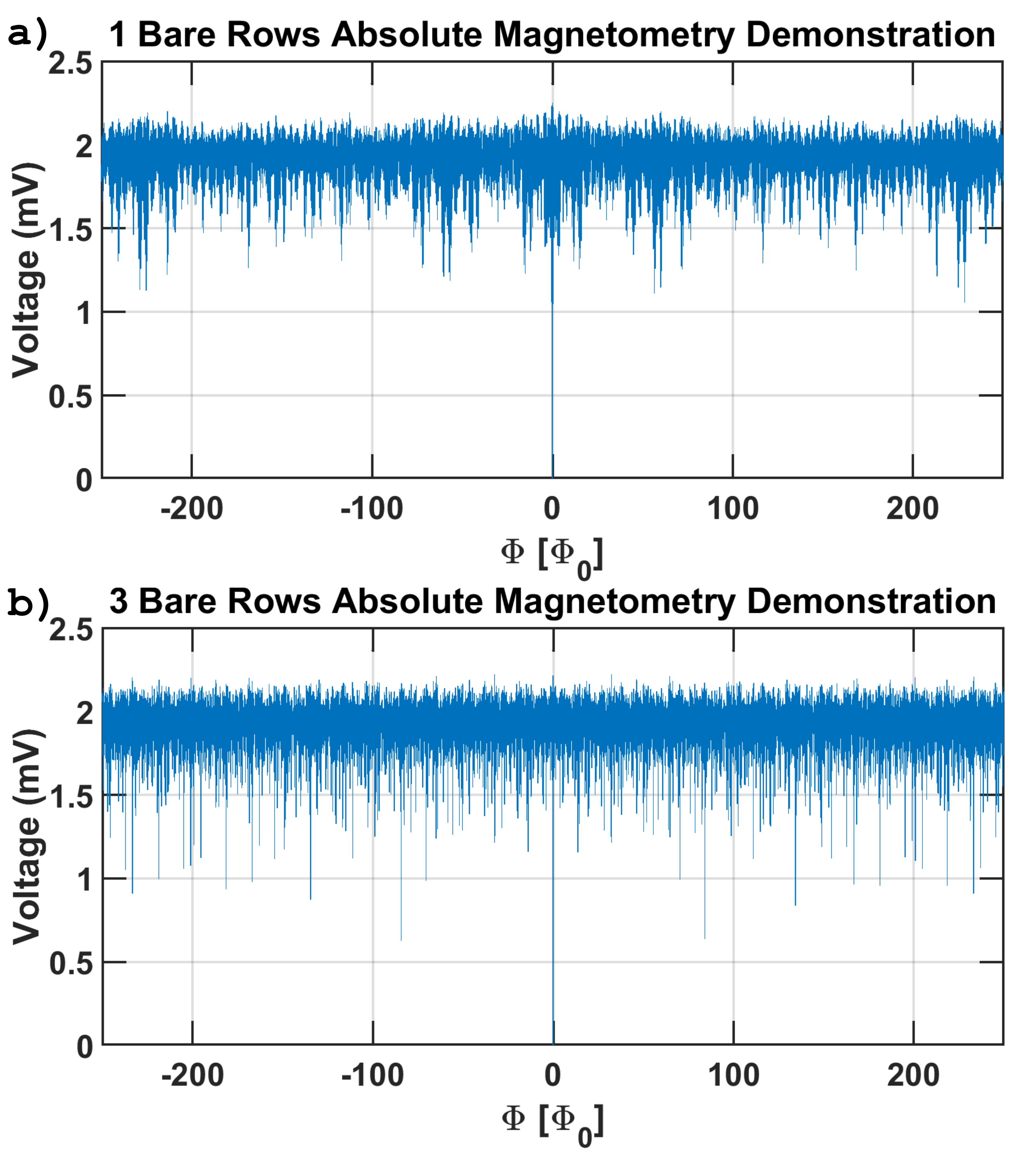}
    \caption{Shown with the blue lines are the Voltage-Magnetic Flux (VMF) response with a wide range of magnetic quantum fluxes (+/- 250) of a) the same devices as in Fig.~3a) in the main section of the paper or b) the same devices as in Fig.~3b) in the main section of the paper. }
    \label{fig::SimData4}
\end{figure}  
\begin{figure}[h]
    \centering
    \includegraphics[width=\linewidth]{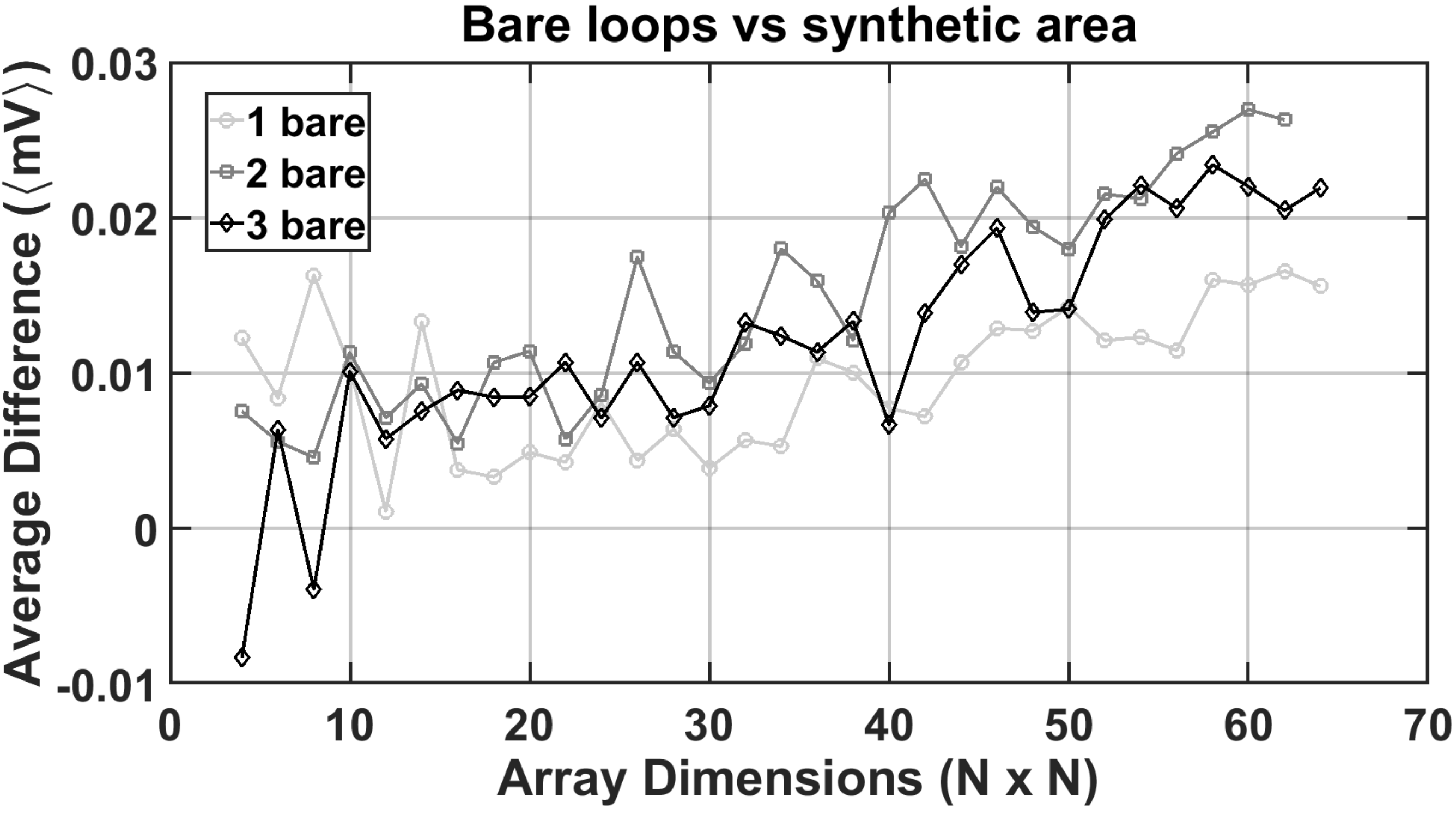}
    \caption{By showing the difference in the theoretical Voltage Response, this figure illustrates the robustness in accuracy of the VMF response for devices with bare loops vs synthetic area, even for $N \times N$ close to $64 \times 64$ and one, two or three lines of bare loops incorporated between lines of DC SQUIDs. These results expand the one of Fig. 3 in the main section of the paper.}
    \label{fig::SimData5}
\end{figure}  

\end{document}